# Temporal Distribution of Solar Cycle 24 Sunspot Groups: Comparison to Cycles 12-23

Jouni Takalo[1]

[1]Space physics and astronomy research unit, University of Oulu, POB 3000, FIN-90014, Oulu, Finland email: jojuta@gmail.com

# Temporal Distribution of Solar Cycle 24 Sunspot Groups: Comparison to Cycles 12-23

Jouni Takalo

Space physics and astronomy research unit, University of Oulu, POB 3000, FIN-90014, Oulu, Finland

email: jojuta@gmail.com

**Abstract** We analyze the temporal distribution of sunspot groups for even and odd cycles in the range SC12-SC24. It seems that cycle 24 is a characteristic even cycle, although with low amplitude. The number of large sunspot groups for cycle 24 is relatively smaller than for the average of both even and odd cycles SC12-SC23, and there is a deep decline of the large groups in the middle of the cycle.

Temporal evolution of the sunspot groups of the even cycles is non-synchronous such that the northern hemisphere distribution of groups maximizes earlier that the southern hemisphere groups. This leads to a double-peak structure for the average even cycle. On the other hand, the distributions of the sunspot groups of odd cycles maximize simultaneously. We show that this double-peak structure intensifies the Gnevyshev gap (GG) for the even cycles, but is not its primary cause. On the contrary, we show that the GG exists for even and odd cycles, and separately on both hemispheres.

We resample all cycles to have equal number of 3945 days and study the difference in the evolution of average total group area and average group area of the even and odd cycles separately. The analysis shows that there is a decline in both total area and average area in the even cycles 1445 days (about four years) after the beginning of the cycle, which is at least 99 % significant for both total and average area. The odd cycles do not have such a clear decline.

# 1. Introduction

Most of the studies of solar cycle 24 have been, thus far, predictions of the size and timing of the cycle. As early as 2005, Svalgaard *et al.* predicted that cycle 24 might be the smallest cycle in 100 years. They predicted the sunspot number maximum to be 75±8 (Pesnell, 2008, this was updated later to 70±2), which is quite near the actual maximum of the monthly mean index in 2014. Pesnell (2008) also gives a review of tens of predictions using different precursors, models and other methods (see Pesnell, 2008 and references therein). The predicted sunspot number index of the maximum varies from under 50 to 180. Especially interesting (according to this study) are the papers by Javaraiah (2007) and Javaraiah (2008). In the papers, the author uses total sunspot group areas and their asymmetry between latitudes 0 - 10 degrees of the preceding cycle as a predictor of the maximum of the next cycle. Javaraiah (2008) concludes that the maximum sunspot index will be 87±7. Choudhuri *et al.* (2007) modelled the last few solar cycles by "feeding" observational data of the Sun's polar magnetic field into their solar dynamo model. Their results fitted to the observed sunspot numbers of cycles 21–23 reasonably well and predicted that cycle 24 will be about 35% weaker than cycle 23, which was extremely near the actual maximum.

There have also been some studies of the cycle 24, although it has been in progress. Watari (2017) studied geomagnetic storms during the ascending phase and maximum of the cycle 24 in 2009-2015. The author stated that there were only seventeen geomagnetic storms with Dst index smaller than -100 nT during this period, and that they showed two-peak structure like the sunspot index itself in years 2011 and 2014.

Sun *et al.* (2015) and Janardhan *et al.* (2018) presented observational analyzes of an unusual polarity reversal during cycle 24. The southern hemisphere reversed polarity in mid-2013 while the reversal in the field in the northern solar hemisphere started as early as June 2012 and was followed by a sustained period of near-zero field strength lasting until the end of 2014 (see also Upton and Hathaway, 2014; Pastor Yabar *et.al.*, 2015).

Joshi and Chandra (2020) used the data of GOES SXR flares from January 2008 to December 2016. The sensors measure solar flares in the 1-8 Å wavelength window, and they are grouped into different classes (B, C, M, X) according to their strength in logarithmic scale. They stated that during the first peak in 2011 there was B-class flares excess activity in the northern hemisphere, whereas C and M class flares excess activity in the southern hemisphere during the second peak (global maximum) of the cycle in 2014. McIntosh *et al.* (2015) found that the peak flare rate occurs at a different time from the sunspot maximum, i.e., about one to two years later. There is, however, smaller maximum during the sunspot maximum and a Gnevyshev Gap (GG) between those flare maxima. This study includes cycles 21-23 and 24 until the end of 2013.

The GG, which is a decline in sunspots and sunspot groups, occurs usually after the maximum of the solar cycle. It is kind of a separatrix between the ascending phase and descending phase of the solar cycle (Gnevyshev, 1967, 1977; Storini *et al.*, 2003; Ahluwalia and Kamide, 2004; Bazilevskaya *et al.*, 2006; Norton and Gallagher, 2010; Du, 2015; Takalo and Mursula, 2018). The time of the Gnevyshev gap is 45-55 months after the start of the nominal cycle length, that is, approximately 33-42 % into the cycle after its start (Takalo and Mursula, 2018).

Kilcik *et al.* (2014) reported that there is a very significant decrease in the number of large sunspot groups in solar cycle 24 compared to cycles 21-23. Furthermore, they found that a significant decrease occurred in the small groups already during solar cycle 23, while no strong changes are seen in the small groups in the current cycle 24 compared to the previous ones. They concluded that the lack of large sunspot groups is mainly responsible for the weak cycle 24. Gopalswamy et al. (2015) found that the number of magnetic clouds (MC) did not decline in cycle 24, although the average sunspot number is known to have declined by ~40%. Despite the large number of MCs, their geoeffectiveness in cycle 24 was very low.
Jiang *et al.* (2015) showed, using simulations, that the weak axial dipole moment (weak polar fields) around the activity minimum of cycle 23 is a result of a number of large low-latitude active regions with "wrong" tilt angle. Since the poloidal magnetic flux related to the axial dipole moment is the dominant source of the toroidal flux that emerges into the subsequent cycle (Cameron and Schüssler, 2015), its low amplitude during the declining phase of cycle 23 caused the weakness of the current cycle 24.

Less attention has been paid to the effects of GG for the geomagnetic activity at the Earth. Richardson and Cane (2012a) and Richardson (2013) have studied the drop of geomagnetic disturbance during the maximum of the solar cycle. They found, especially, a tendency for the Kp-index storm rate to fall during fourth year of the cycle. They also associated several occasions of decline in geomagnetic activity aa-index around solar maximum to the Gnevyshev gap during solar cycles 20 - 24 (Richardson and Cane, 2012b).
Takalo (2020b) has also shown that there is a GG related decline in near-Earth IMF and subsequently in the geomagnetic disturbances at the Earth as measured with Ap- and Dst-index.

In this study, we mainly analyze the temporal distribution of sunspot groups of Solar Cycles (SC) 12-24. Special attention is paid to the differences and similarities of cycle 24 compared to other even cycles. This article is organized as follows. Section 2 presents the used data. In Section 3 we concentrate on the temporal distributions of sunspot groups for cycle 24 and for even and odd solar cycles SC12-SC23. In Section 4 we study temporal evolution of the sunspot group area in three categories. In Section 5 we analyze the shape of cycle 24 and compare it to SC12-SC23, and give our conclusions in Section 6.

## 2. Data and Methods

2.1 Sunspot Group Data

For the analysis of sunspot cycles 12-24 we use recently published catalogue of sunspot groups by Mandal *et al.* (2020) (later M set). In Takalo (2020a) we used sunspot group data for SC8-SC23 by Leussu *et al.* (2017) (L set) to study the latitudinal distribution of sunspot groups. We use this data set here as a reference set, although it does not contain the area information of the sunspot groups. The main (common) data for both of these sets are the Royal Greenwich observatory (RGO) measurements between 1874-1976. The biggest difference of these data are the years after 1976, i.e. solar cycles SC21-SC23 (and the first half of cycle 24). In L set these years are constructed from SOON (Solar Observatory Optical Network) data for the period 1976-2015 (Neidig *et al.*, 1998). In M set the composite data sets from Kislovodsk, Pulkovo and Debrecen observatories were used for the years from 1976 to October 2019, i.e., containing (almost) solar cycle 24. In addition, the number of groups is much smaller in L set, because it contains only the first appearance of the sunspot group in its record, while M set contains daily appearance of the groups. Thus, the lifetime of each group also affects the number of the groups. Figure 1 shows butterfly patterns for both data sets. L set is shown with blue dots for SC12-SC24 (SC24 until the end of 2015) and M set with red dots for SC12-SC23 and SC24 until October 2019. The sets are practically the same for SC14-SC20, but differ for SC21-SC23. The main reason for the difference is that L set has only integer values for the latitudes during these cycles, but different data sources bring also some deflection to the points of the sunspot groups. Some extra points in the SC12 and SC13 for L set are due to the refilling of the RGO group data with the observations of G. Spörer until 1894 (Diercke *et al.*, 2015). Note also, that there are some groups at high latitudes during 2018-2019 for M set. Some of these groups probably belong already to the cycle 25 (Phillips, 2019).

**3. Temporal distribution of sunspot groups for SC12 - SC24**

In order to compare the temporal evolution of the cycle 24 to the even and odd SC12-SC23, we normalize the time axis simply calculating time as $T_{new} = (T_i - T_{min})/L$, where $T_i$ is the original time of each group, $T_{min}$ is the time for the leading minimum of the cycle, and $L$ is the length of the cycle. The normalized time is thus $0 \leq T_{new} \leq 1$. The dates of the cycle minima and the cycle lengths for SC12-SC24 used in this study are shown in Table1 (Takalo and Mursula, 2020).

Figure 2a shows the sunspot groups for northern (blue) and southern hemisphere (red) of the solar cycle 24. The time consists of 130 points (months), which is the average length of the cycles SC12-SC24. Here and elsewhere (if not otherwise mentioned) we use moving average smoothing through seven points such that the end points of the window have only half of the weight from the other points (this is called trapezoidal smoothing). Time axis of the cycle is shown as both years and normalized time. It is evident that the maximum in the northern hemisphere occurs earlier than the maximum in the southern hemisphere (the separation of the highest peaks is about two years). Furthermore, northern groups have flatter evolution after the maximum and a side maximum before the descending phase of the cycle. The southern groups have also a side maximum about a year before the highest maximum and a gap between these two maxima. Note that the gap for northern hemisphere (between the peaks) exists also about year earlier than the southern gap. Figure 2b depicts the latitude distribution of the sunspot groups for cycle 24. The northern hemisphere has more sunspot groups (7847) than southern hemisphere (6888). It seems also that the excess of sunspot groups in the north are due to small groups, i.e., groups with area ≤100 µHem (millionths of hemisphere) and they occur mostly in the descending phase of the cycle. There are 4753 and 3948 small groups (≤100 µHem) in the northern and southern hemisphere, respectively. On the contrary, there are 475 and 516 large groups (≥500 µHem) in the northern and southern hemisphere, respectively.

Figures 3a and 3b depict the average distributions of northern and southern hemisphere sunspot groups for even and odd cycles of SC12-SC23, respectively, and total number of groups. Note that for the odd cycles both hemispheres are simultaneous, while the northern groups maximize earlier than the southern groups for even cycles. Furthermore, the groups in the even cycles have flatter and wider structure in the distribution at its maximum than in the odd cycles, which has a sharp maximum in total number of groups. Note also that there is a gap, GG, after the first and the second maximum of the total count for even cycles, which lasts more than a year. The second peak in the even cycles exists mainly in southern hemisphere, i.e., the double peak for even cycles is caused partly by non-synchronous maxima in the northern and southern hemispheres of the sun. The highest peaks are located at about 34 % and 48 % from the start of the cycle for northern and southern hemisphere groups, respectively.

In this respect, cycle 24 is a characteristic even cycle, but the gap between northern and southern hemisphere maxima is even wider than the average of even cycles between SC12-SC23. (It should, however, be noted that statistics for one low cycle is worse than the statistics for six even and on the average much higher cycles.) Figures 3c and 3d show the average group distributions of even and odd cycles for northern and southern hemisphere of SC12-SC23, respectively. Note that in the northern hemisphere even cycles have maximum earlier than odd cycles, but in the southern hemisphere vice versa. The odd cycles have maximum at 35 % from the start of the cycle in both hemispheres, i.e., they are synchronous with each other.

Figure 4 shows the temporal distribution of the sunspot groups for the solar cycle 24 with absolute latitude greater than 15 and absolute latitude smaller or equal to 15 degrees. We use 15 degrees as divider in different clusters of groups, because it seems to be kind of separatrix for ascending and descending phases of the cycles (Takalo, 2020a). The sunspot groups with absolute latitude over 15 degrees concentrate evidently to the first half of the cycle and sunspot groups with absolute latitude smaller than 15 degrees to the second half of the cycle. There is a trough at about 0.35 of normalized time. This corresponds the first half of the year 2013 in real time-line, i.e., starting about four years from the beginning of the cycle. Interestingly figure 4 resembles very much figure 2a. This means that northern hemisphere groups dominate the higher latitudes, while southern hemisphere groups dominate the lower latitudes, except the last couple of years of the descending phase. Note that we have divided the cycles using the minima as starting and ending points of the cycle (see Table 1.). That is why there may be low latitude groups, which belong to the previous cycle, in the start of the cycle. Similarly, the last high-level groups may belong already to the next cycle. This, however, does not affect to the overall shape of the sunspot group distributions.

Figures 5a and 5b depict distributions of sunspot groups of both hemispheres for all groups (black) absolute latitude greater than 15 degrees (blue) and smaller or equal to 15 degrees (red) for M set and L set of even cycles SC12-SC23, respectively. The troughs are in both sets at about 0.4 of the normalized time for M set and 0.35 for L set from the start of the cycles. The reason for the difference of the trough is that M set includes the whole lifetime of the cycle while L set contains only the first appearance of the emerging group. We believe the trough is the point for Gnevyshev gap (GG) during the solar cycle. We also believe that this is the first time the GG is detected as a trough in the latitude distribution data between these two categories of sunspot groups.

In comparison, we show the distributions of sunspot groups of both hemispheres for all groups located at absolute latitude greater and smaller than 15 degrees for M set and L set of odd cycles SC12-SC23 in Figs. 6a and 6b, respectively. Note that the sunspot groups of higher latitudes than 15 degrees and lower latitudes than 15 degrees do not have so clear v-shaped trough as the even cycles between their maxima. There is a trough in the sum of the distributions of L set during the maximum, but the decline is smaller and narrower than for even cycles. The M set for odd cycles do not have trough but a sudden decline after the maximum in the smoothed sum curve. This is probably because we have the whole lifetime of the sunspot groups in the M set and this flattens the fluctuations in the distribution. This is also the reason why the drop (related to GG) during the maximum in figure 5a for M set is not as steep as the drop of the L set in figure 5b. The GG exists only as a starting point of the descending phase after the maximum for the M set odd cycles.

## 4. Size evolution of sunspot groups for SC12-SC24

Figure 7 shows the temporal distribution of the number of sunspot groups for the solar cycle 24 with areas smaller (or equal) than 100 μHem (millionths of the hemisphere), between 100 and 500 μHem, and larger (or equal) than 500 μHem. In the beginning and at the end of the cycle there are only small sunspot groups. Especially, small sunspot groups (≤100 μHem) dominate the latest third of the cycle. It is evident that the smaller the area of the group the wider is their distribution. Note also a couple years decline in all groups after 2012.

Figure 8a and 8b show similar distribution, but separately for even and odd solar cycles 12-23, respectively. The striking difference between the even and odd cycle area distributions is that odd cycles have narrower maximum for the small and medium sized areas. The average even cycle is flatter and has a double peaked maximum for the small and medium sized areas. There is a decline at about 0.35-0.4 in both even and odd cycles for largest area groups, but a little deeper for the even cycles. The cycle 24 reminds the average even cycle, but due to poorer statistics, its distribution has structure that is more rugged. The cycle 24 has, however, relatively fewer large sunspot groups (area ≥500 μHem) than the cycles between SC12-SC23 (Kilcik *et al*., 2014). We calculated that cycle 24 has 6.5 % (1.6 %) groups with area ≥500 μHem (area≥1000 μHem) from total of 15184 sunspot groups. The average even cycle between SC12-SC23 has 9.1% (2.3%), and the average odd cycle 9.5% (2.5%) with area ≥500 µHem (area≥1000 µHem) from total of 17542 and 20618 sunspot groups, respectively.

Takalo and Mursula (2020) showed, using different data set, that the GG is pronounced for the large sunspot groups. Figure 9a and 9b show the distributions of sunspot groups with area≥1000 µHem for even and odd cycles between SC12-SC23, respectively. The red curve shows the evolution of the average latitude of the emerging groups as calculated from the L set. Note that for the even cycles the GG is deep and four bins wide (here one bin is about 1.5 months), which means about half a year. Note also that the trough is located at the time when average of the group is about 15 degrees of latitude. On the other hand, the GG for odd cycles is just one bin wide, although located in the same 15 degree crossing point with the average latitude curve. Figure 9c shows the sunspot groups with area≥1000 µHem for cycle SC24. For statistical reasons (too few large groups), we use here still wider bin, which is about three months. It is, however, evident that there is a huge gap at the site when the average curve crosses the 15 degrees of the latitude.

Kilcik *et al*. (2011) found that large sunspot groups appear to reach their maximum in the middle of the solar cycle, while the small SG numbers generally peak much earlier. Our analysis shows that this is true for the cycle 24, but not common for the average even cycle. Actually, the Figs. 8a and 9a show that there are slightly more large sunspot groups in the first maximum (before GG), and more small sunspot groups in the latter maximum for average even cycle. In addition, both small and large groups are slightly more abundant in the foreground of the mainly single-peaked maximum for average odd cycle. Note, however, that Kilcik *et al*. (2011) used Zürich classification (McIntosh, 1990) for the sunspot groups, which is not in use in our analysis. Norton and Gallagher (2010) found that there was phase difference from zero to fourteen months in the maximum between hemispheres in the sunspot number and sunspot area data for cycles 12 - 23. They, however, concluded that the double peak and the GG are not caused by the different timing of the maxima in the northern and southern hemispheres of the sun. This is actually also my opinion. The double peak in the even cycles is caused from different timing of the maxima in different hemispheres (on the average), but the GG is a separate phenomenon and exist in both even and odd cycles. It is evident from Fig. 10 that the GG exists in both hemispheres, interestingly here for southern hemisphere more clearly. In addition, the GG exists also in the odd cycles, although odd cycles are maximizing quite simultaneously in both hemispheres. It is, however, possible that the double-peaked structure intensifies the GG phenomenon for the even solar cycles.

## 5. The shape of the cycle 24 compared to the cycles SC12-SC23

In order to study the shape of the cycles using sunspot groups, we resampled daily sunspot group areas and daily sunspot group numbers such that all cycles have the same length of 3945 time steps (days), which corresponds to the average 10.8 years of the cycles.

Figure 11 depicts the average daily total area of sunspot groups for even (blue) and odd (red) cycles of SC12-SC23. Note that the green curve shows average area when cycle 24 is included to the even cycle curve. We employ here also trapezoidal smoothing. The smoothing window here is 61 days (two months).

Evidently, there is a clear decline in the total areas between the first two vertical dotted lines. This corresponds to the interval between 1445-1576 days from the beginning of the minimum, i.e., 37 % of the average cycle length, constituting four months between the lines. (Notice that this is in the halfway between the northern and southern maxima in the distributions of the sunspot groups of the even cycles.) There is an extended decline until about 1725 days (marked with thinner dashed line in the Fig. 11) but this decline exist only in some even cycles, e.g., in cycle 24. The four-month decline, which we believe to be related to GG, is statistically very significant according to two-sided T-test for unequal means (see the T-test method in Takalo, 2020a and Takalo and Mursula, 2020). The mean value for the total area of the interval 1445-1567 is 1342 µHem, while for the earlier four months it is 1610 µHem, and for the following four months 1478 µHem (The horizontal red lines show these mean values in the figure. Notice also that the latter interval is inside the second decline region). The mean value for the average group area of the interval 1445-1567 is 172.3 μHem (7.8 groups/day), while for the earlier four months it is 200.4 μHem, and for the following four months 197.0 μHem. When comparing the interval 1445-1567 to four months earlier and four months later both the average total area and the average group area have 99 % significance with p-values $<5.5\times10^{-10}$ and $<1.9\times10^{-7}$, respectively, when calculated from daily raw (unsmoothed) data using two-sided T-test. The mean value of total group area for the cycle 24 in the interval 1445-1567 is only 909 μHem, while for the four months earlier it is 1235 μHem. The group area stays low for exceptionally long period of 16 months before the second and highest maximum begins in the beginning of 2014.

Note that for the interval 1445-1567 of the odd cycles the average total area is 2089 μHem, and the average group area 208.9 μHem (10.0 groups/day). Therefore, there are over 20 % more groups in odd cycles than in even cycles during this interval. Furthermore, this interval does not differ significantly from the preceding or following four-month intervals for odd cycles.

## 6. Conclusion and Discussion

We have analyzed temporal distribution of sunspot groups for the even and odd cycles in the range SC12-SC24. Special attention has been paid to the cycle 24 and its comparison to previous even and odd cycles. It seems that cycle 24 is a characteristic even cycle, although with low amplitude and longer than average decline between the two maxima. Similarly to other even cycles the northern hemisphere sunspot groups maximize earlier than the southern hemisphere groups. The first (smaller) maximum of the cycle 24 at the end of 2011 seems to be due to northern hemisphere groups, while the second greater maximum in 2014 to southern hemisphere groups. The northern hemisphere of the cycle is dominating in the number of sunspot groups, but mainly in small (area≤ 100μHem) sunspot groups. Furthermore, the excess of small groups in the northern hemisphere exist in the descending phase of the cycle. This may be due to above-mentioned ambiguous northern polar field reversal (Sun *et al*., 2015, Janardhan *et al*., 2018). On the other hand, the number of large sunspot groups (area ≥500 μHem) is relatively smaller than the average of both even and odd cycles SC12-SC23. In addition, the southern hemisphere has more large groups than the northern hemisphere of the cycle 24.

Temporal evolution of the sunspot groups of even cycles is non-synchronous such that the northern hemisphere distribution of groups maximizes earlier that the southern hemisphere groups. This leads to a double-peaked structure for the average even cycle. On the other hand, the distributions of the sunspot groups of odd cycles maximize simultaneously, and their average maximum is between the two maxima of the even cycles as plotted in the normalized time. There is, however, a recent tendency that the two last odd cycles are also double-peaked, while the odd cycles 13, 15, 17 and 19 have only one maximum.

We have also divided sunspot groups in two clusters according to latitude: absolute latitude greater than 15 degrees and absolute latitude smaller or equal than 15 degrees. The temporal distribution of these clusters for cycle 24 are similar to earlier even cycles such that there is a v-shaped trough between the maxima of the distributions. This leads to a clear gap at the time, when the sunspot groups cross (on the average) the 15 degrees of the heliographic latitude. At the same time, toroidal field generates equatorward migrating sunspots and poloidal field is drifted towards the polar regions, which leads to a field reversal at polar caps (Dikpati and Charbonneau, 1999). We believe that this is the right timing of the GG phenomenon. It should, however, be emphasized that GG and double peak structure are different phenomena. The GG occurs in both even and odd cycles, and separately in both hemispheres during the reversal of the magnetic field at the polar region in the corresponding hemisphere. We have shown that the difference in the northern and southern sunspot group maxima for the even cycles may pronounce the GG for the even cycles. This is validated by the result, that reversals of polar magnetic field in Solar Cycles 21, 22, 23, and 24 were completed first in the northern hemisphere, and 0.6, 1.1, 0.7, and 0.9 years later in the southern hemisphere, respectively (Pishkalo, 2019). Note that the difference is larger for the even cycle 22 and 24 than for odd cycles 21 and 23, (although the cycle 24 reversal was somewhat unusual).

We have resampled all cycles to have equal number of 3945 days (10.8 years) and studied the difference in the evolution of average total group area and average group area of the even and odd cycles separately. The analysis shows that there is a decline in both total area and average area in the even cycles 1445 days after the beginning of the cycle. This last about 120 days (four months) and is statistically at least 99 % significant for both total and average area. We believe that this is related to the more pronounced GG of the even cycles. The odd cycles do not have such a clear decline. This phenomenon is so clear that dynamical models of the Sun should consider this. The basic reason for the GG is not the scope of this study, and it is an open question now, but some good attempts have been presented (Georgieva, 2011; Karak *at al*., 2018). There have also been suggestions about a relic field in the Sun (Bravo and Stewart, 1996; Mursula *et al*., 2002; Song and Wang, 2005). Because field reversal occurs during (or soon after) the maximum of the cycle, the relic field could also affect to the different behavior of the sunspots of the even and odd cycles during or after the maximum.

Figure 12 shows the importance of the GG for the geomagnetic disturbances at the Earth. It depicts the sequence of the GG phenomenon from the sun through IMF to the Earth for the even cycles 20, 22, and 24. Note that there are two drops in the GG interval for the group areas: first between 1445-1567 and second between 1567-1725. The first drop has only a smaller response in IMF B and Ap-index, but the second one has a striking decline in both IMF B and geomagnetic activity. Both the magnetic field intensity B and Ap-index reach by far the smallest values at the end of this interval (except of course at the beginning and end of the cycle). Note that Ap-minimum exists a little later than IMF B minimum (Takalo, 2020b).

**Acknowledgements**

The sunspot group area and group number data were downloaded from http://www2.mps.mpg.de /projects/sun-climate/data.html. The sunspot group data used as a reference were retrieved from VizieR database (https://vizier.u-strasbg.fr/viz-bin/VizieR?-source=J/A+A/599/A131\&-to=3). The dates of cycle minima were obtained from the National Geophysical Data Center (NGDC), Boulder, Colorado, USA (ftp.ngdc.noaa.gov).

**Table 1.** Sunspot cycle lengths (in years) and dates (fractional years, and year and month) of (starting) minima for cycles 12-24 (NGDC, 2013).

| Sunspot cycle number | Fractional year of minimum | Year and month of minimum | Cycle length |
|---|---|---|---|
| 12 | 1879.0 | 1878 December | 10.6 |
| 13 | 1889.6 | 1889 August | 12.1 |
| 14 | 1901.7 | 1901 September | 11.8 |
| 15 | 1913.5 | 1913 July | 10.1 |
| 16 | 1923.6 | 1923 August | 10.1 |
| 17 | 1933.7 | 1933 September | 10.4 |
| 18 | 1944.1 | 1944 February | 10.2 |
| 19 | 1954.3 | 1954 April | 10.5 |
| 20 | 1964.8 | 1964 October | 11.7 |
| 21 | 1976.5 | 1976 June | 10.2 |
| 22 | 1986.7 | 1986 September | 10.1 |
| 23 | 1996.8 | 1996 October | 12.2 |
| 24 | 2009.0 | 2008 December | 11.0 |
| 25 | 2020.0 | 2019 December | |

Figure captions.

**Figure 1.** Butterfly patterns of the sunspot groups by Leussu et al. (2017) (L set, blue dots) and by Mandal et al. (2020) (M set, red dots).

**Figure 2.** Distribution of sunspot groups of northern and southern hemisphere for cycle 24. Time axis is shown both in years and in normalized time. b) The latitude distributions of three categories of sunspot groups of cycle 24 for the southern and northern hemispheres. (Bin size is about one degree in latitude).

**Figure 3.** Panels a) and b) Distribution of sunspot groups of NH, SH and total for even cycles and odd cycles between SC12-SC23, respectively. Panels c) and d) Distributions of sunspot groups of even and odd cycles for northern and southern hemispheres, respectively.

**Figure 4.** Distribution of sunspot groups of both hemispheres for absolute latitude 15 greater than 15 degrees (blue) and smaller or equal to 15 degrees (red) and total (black). Time axis is shown both in years and in normalized time.

**Figure 5.** Distribution of sunspot groups of both hemispheres for all groups, absolute latitude greater than 15 degrees and smaller or equal to 15 degrees a) for M set of even cycles SC12-SC23, and b) for L set of even cycles SC12-SC23.

**Figure 6.** Similar to Fig. 5, but now for odd cycles SC12-SC23.

**Figure 7.** Distribution of sunspot groups of cycle 24 for areas smaller (or equal) than 100 µHem (millionths of the hemisphere), between 100 and 500 µHem and larger (or equal) than 500 µHem.

**Figure 8.** Distribution of sunspot groups for areas smaller (or equal) than 100 µHem (millionths of the hemisphere), between 100 and 500 µHem and larger (or equal) to 500 µHem a) for even cycle SC12-SC23, b) for odd cycles SC12-SC23.

**Figure 9.** Distribution of sunspot groups of both hemispheres for all groups with area≥1000 µHem a) for even cycles SC12-SC23, b) for even cycles SC12-SC23, and c) for cycle SC24. The red curve is the mean latitude of emerging groups throughout the average solar cycle SC12-SC23.

**Figure 10.** Distribution of sunspot groups of a) for northern hemisphere, b) for southern hemisphere of groups with area≥1000 µHem of even cycles. The magenta curve is the mean latitude of emerging groups throughout the mean solar cycle SC12-SC23.

**Figure 11.** Average daily total area of sunspot groups for the even and odd cycles SC12-SC24. The vertical dashed lines show the clear change points in the daily group areas for even cycles. The horizontal red lines show the mean values of four-month data around the interval 1445-1567 days.

**Figure 12.** Top panel: The average group area for the even cycles 20, 22 and 24. Middle panel: The average IMF |B|-component for the same cycles. Bottom panel: The average Ap-index for the same cycles.

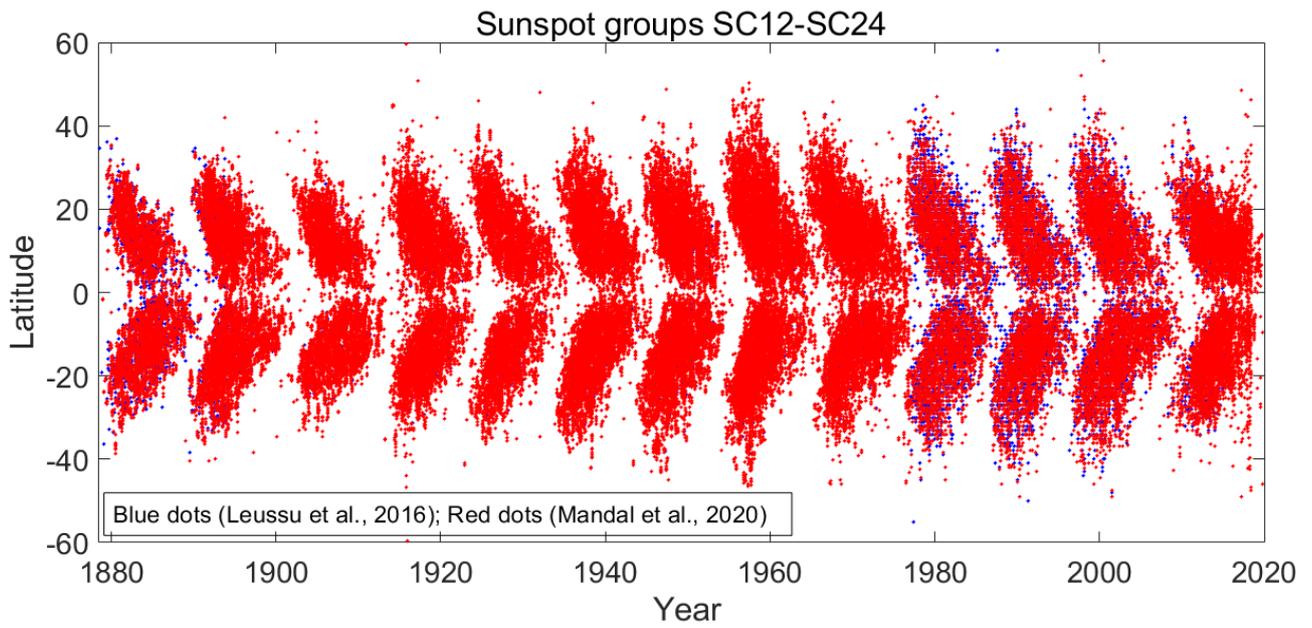

**Figure 1.**

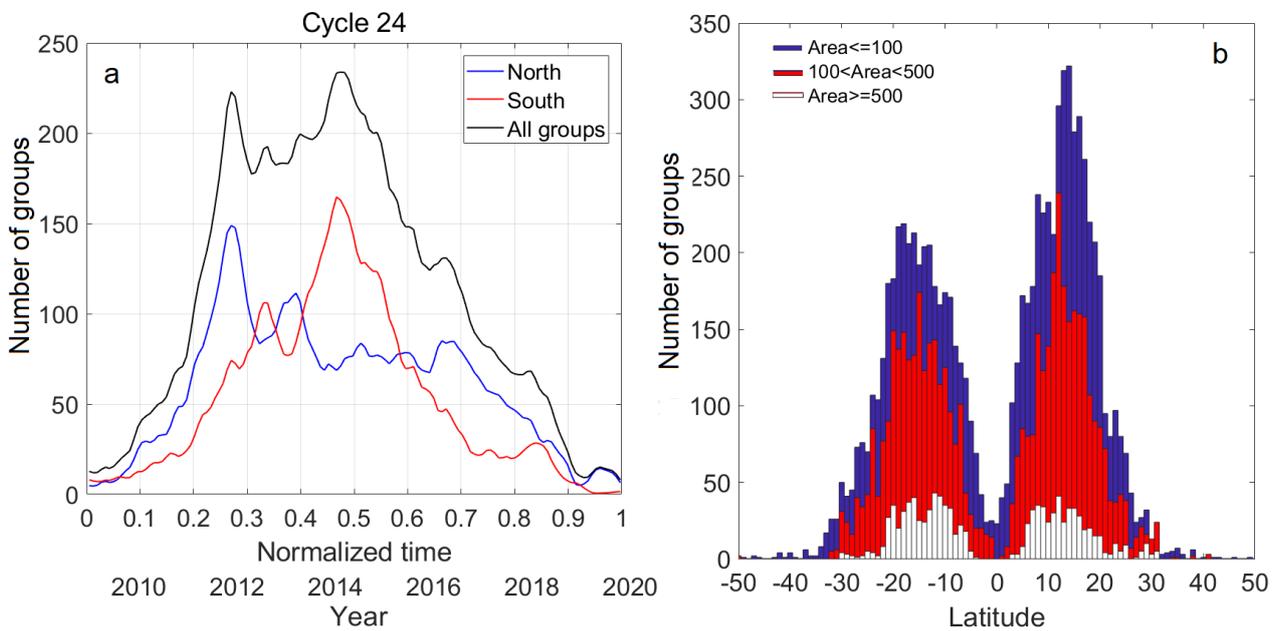

**Figure 2.**

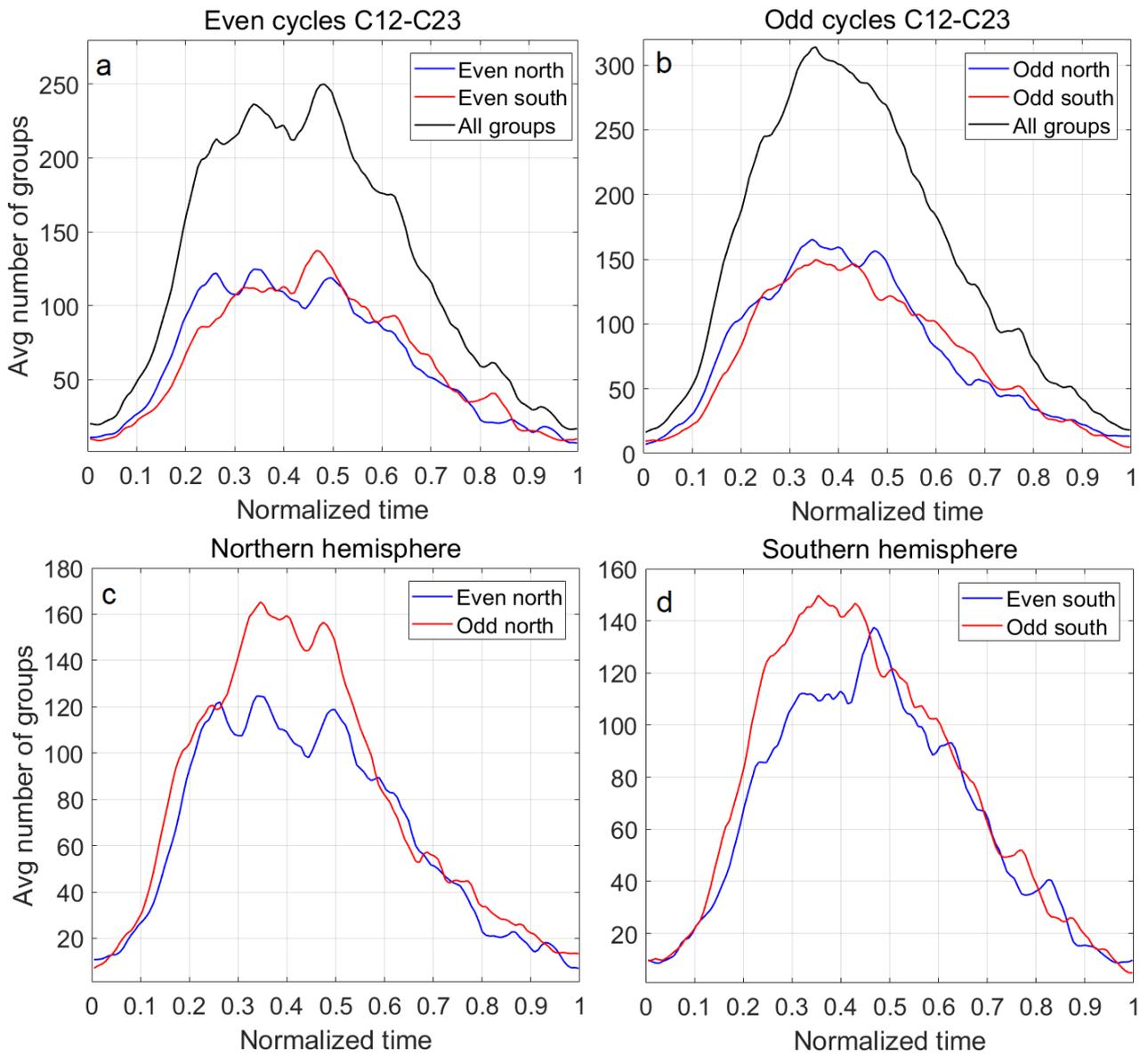

**Figure 3.**

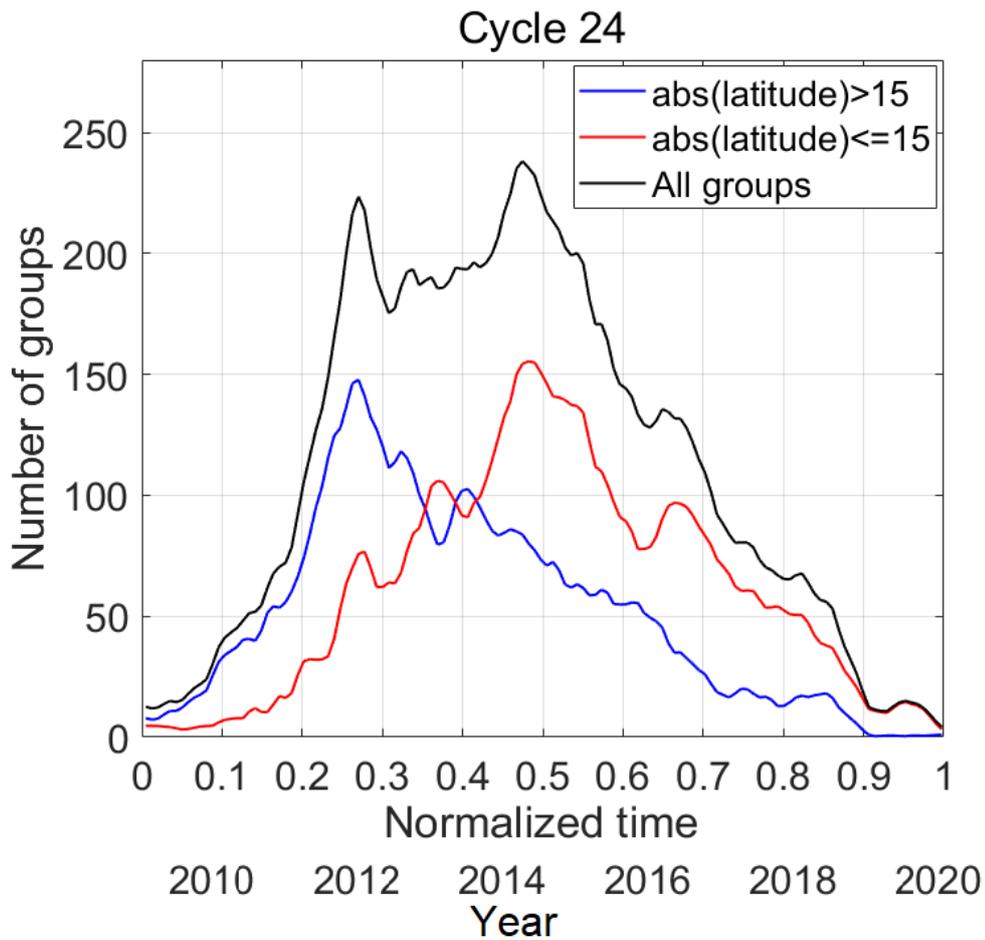

**Figure 4.**

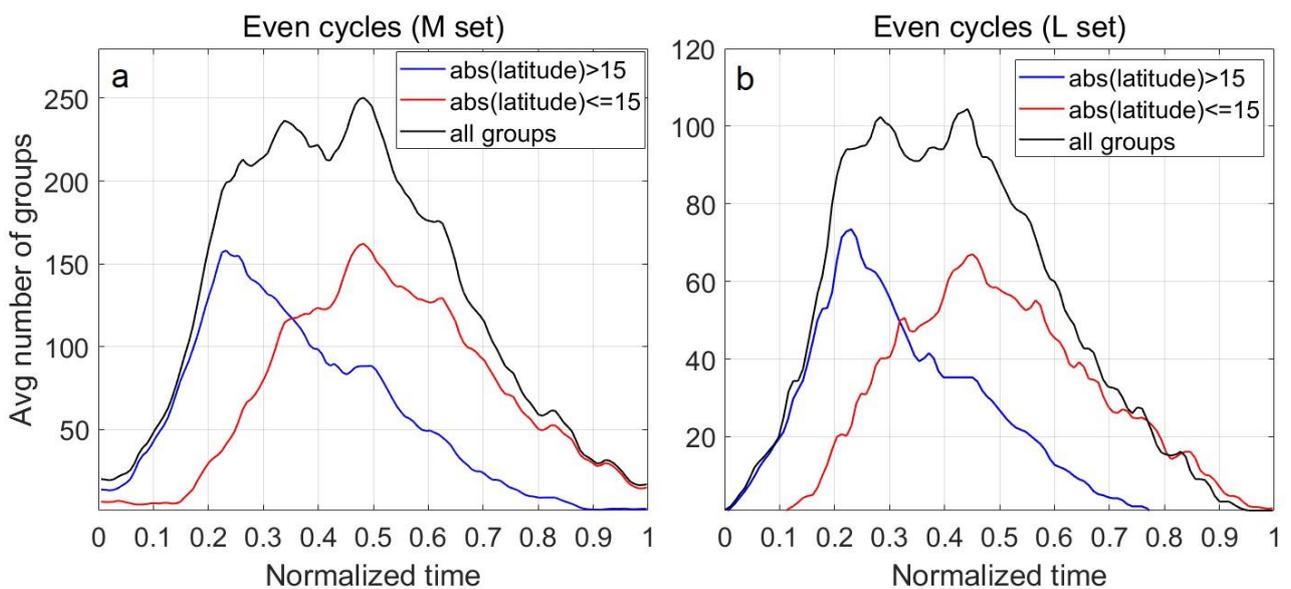

**Figure 5**

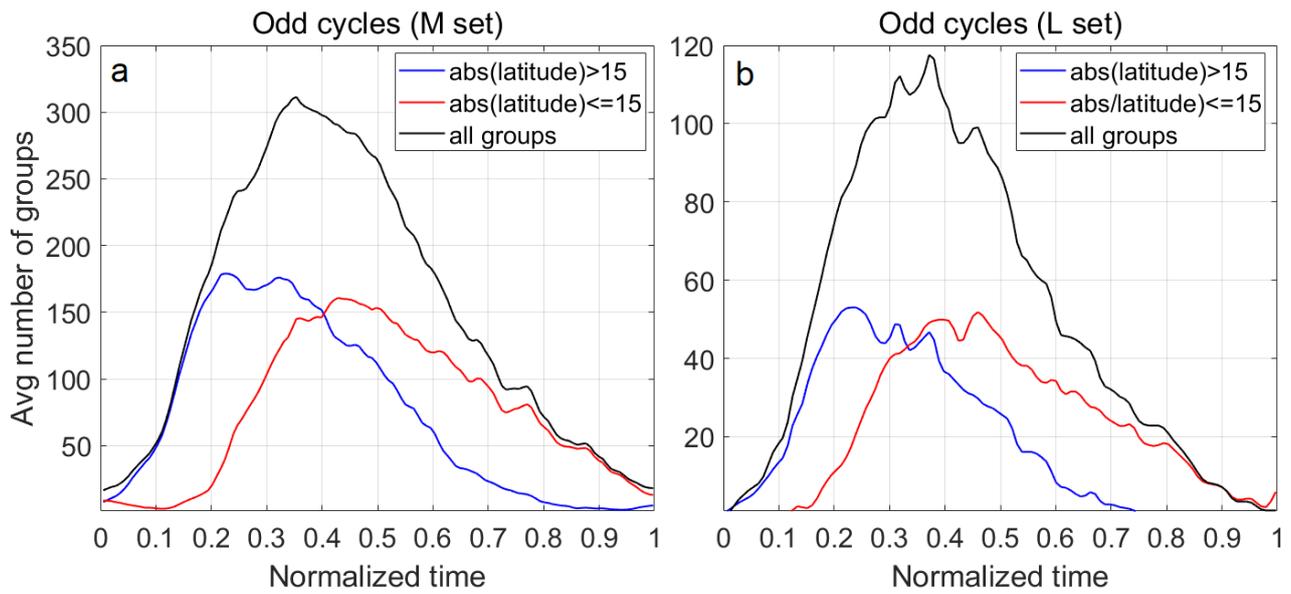

**Figure 6.**

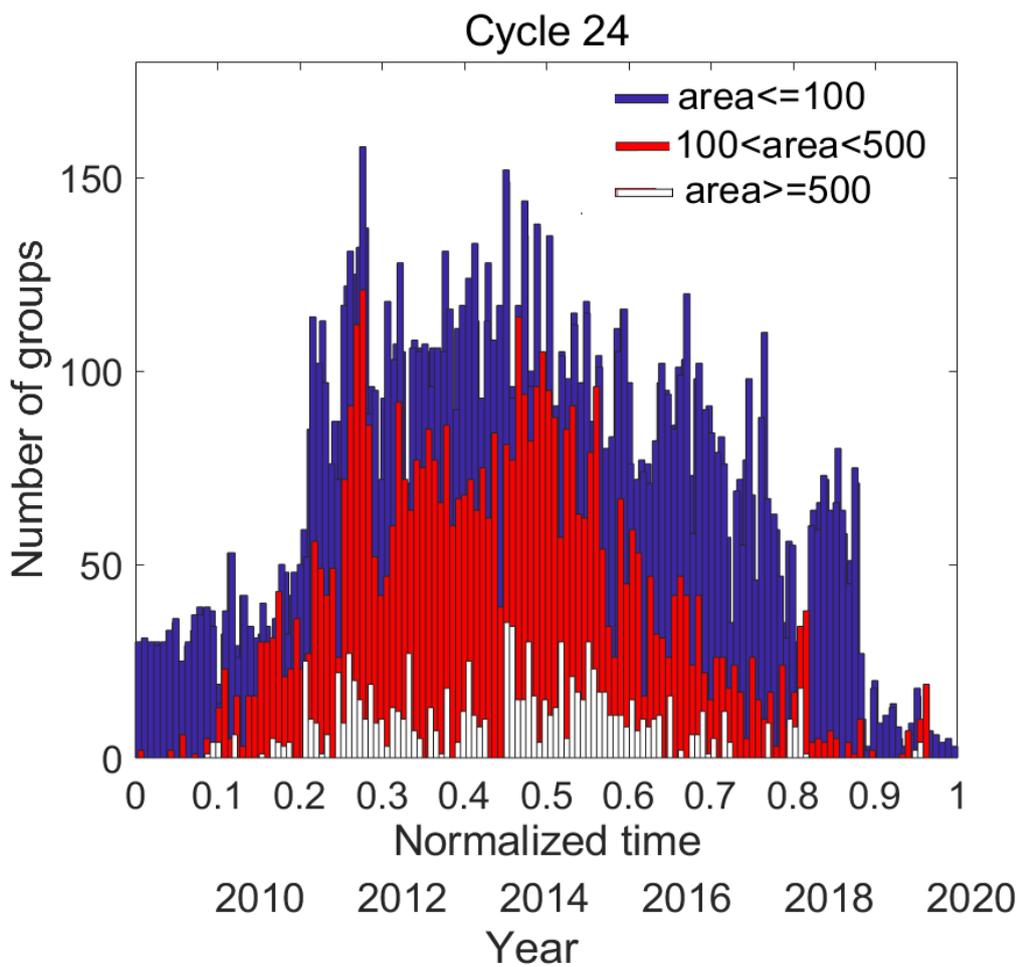

**Figure 7.**

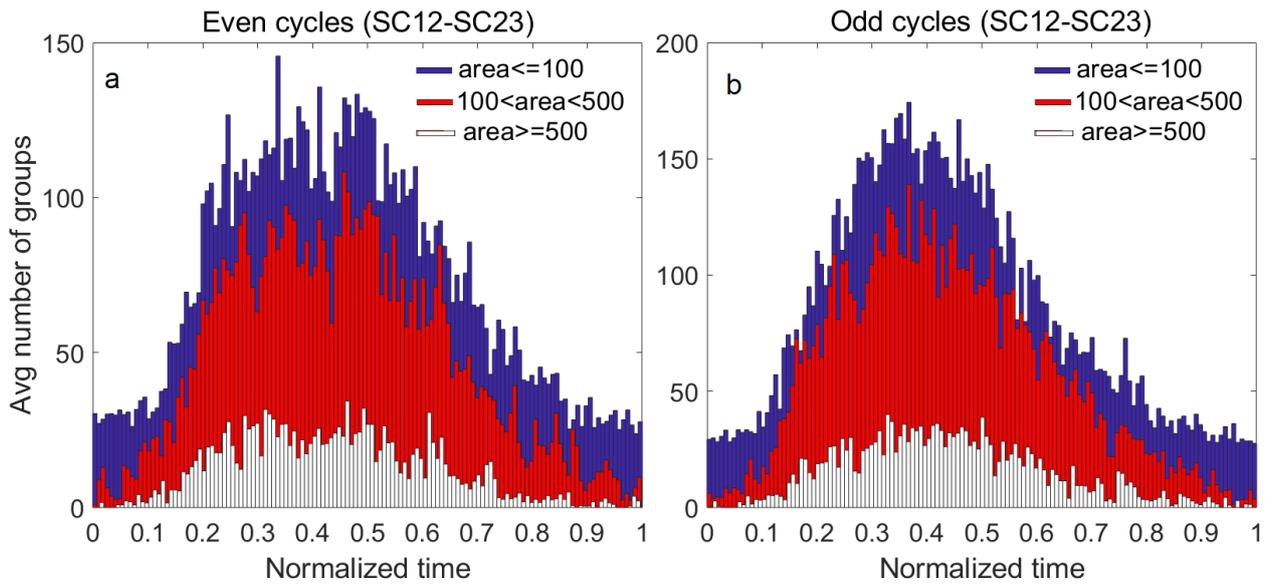

**Figure 8.**

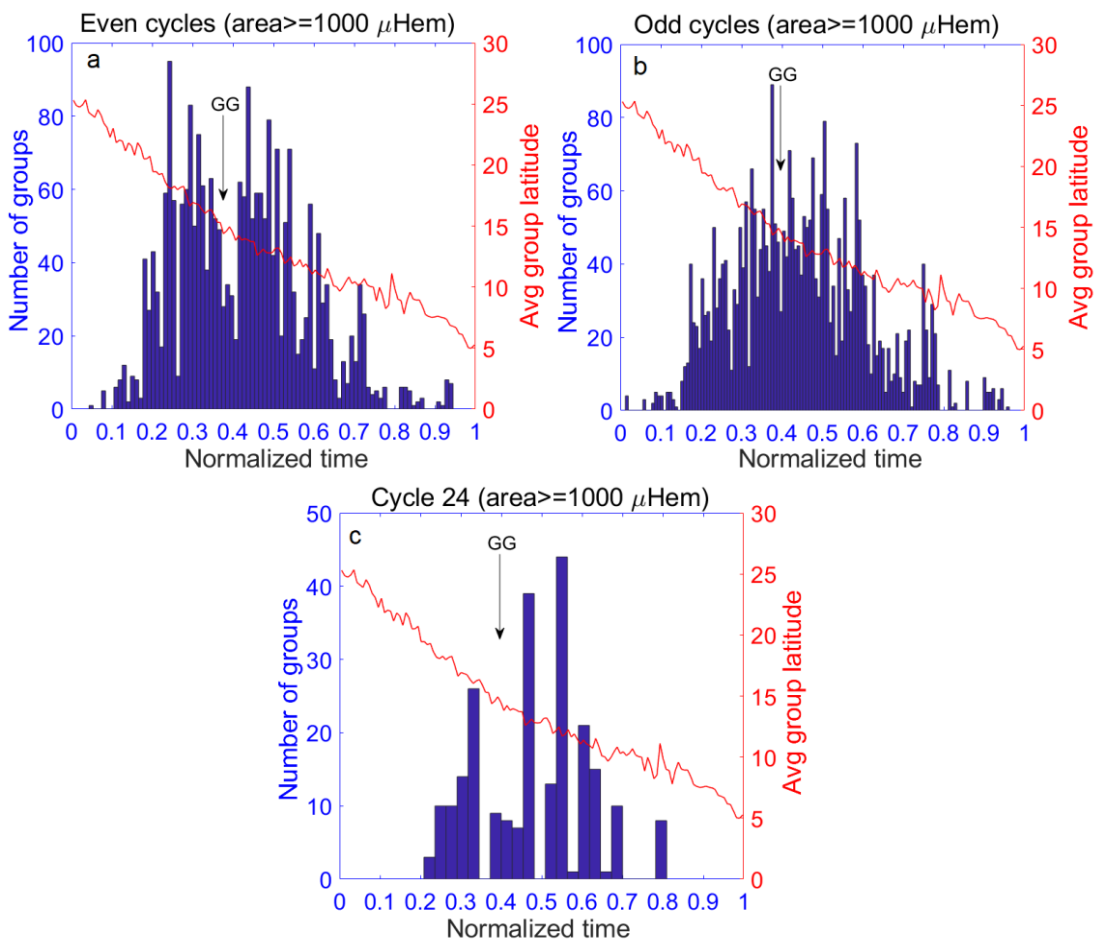

**Figure 9.**

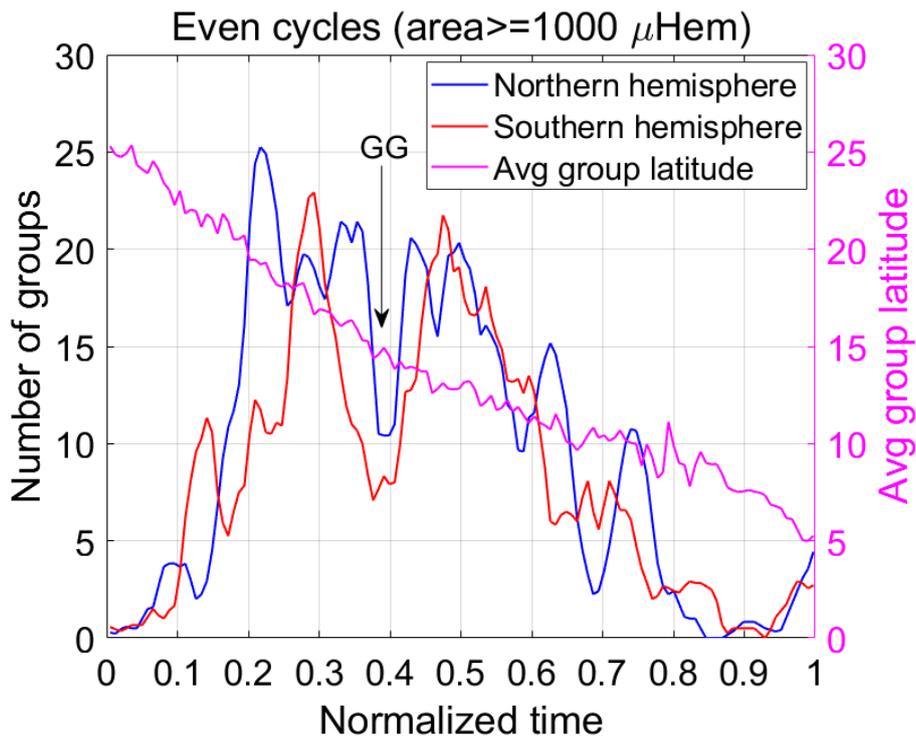

**Figure 10.**

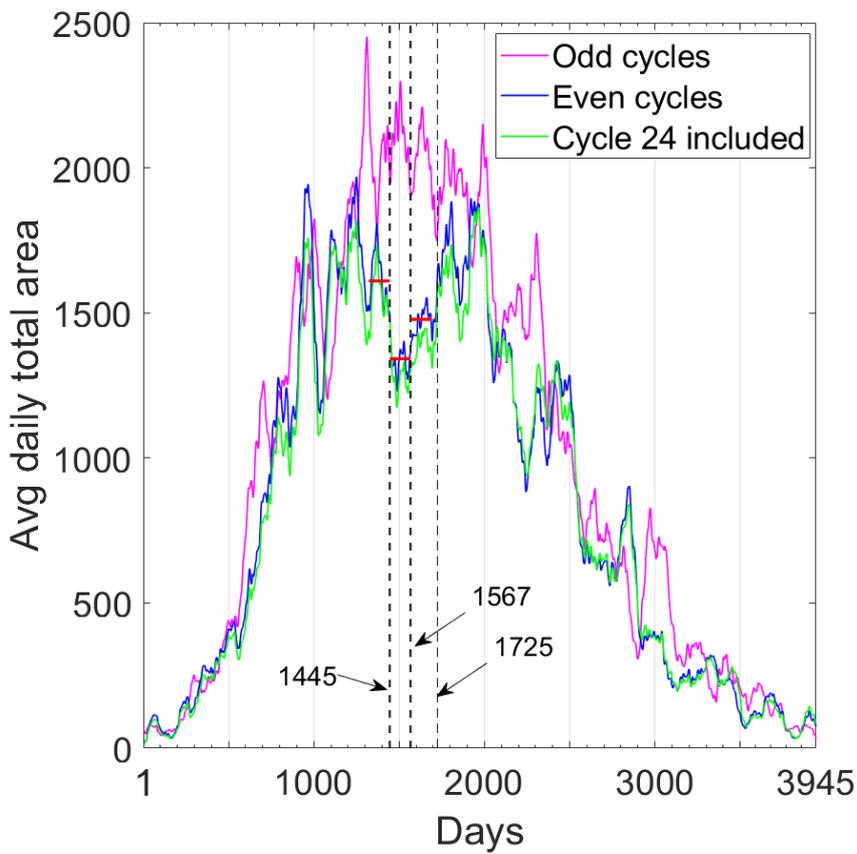

**Figure 11.**

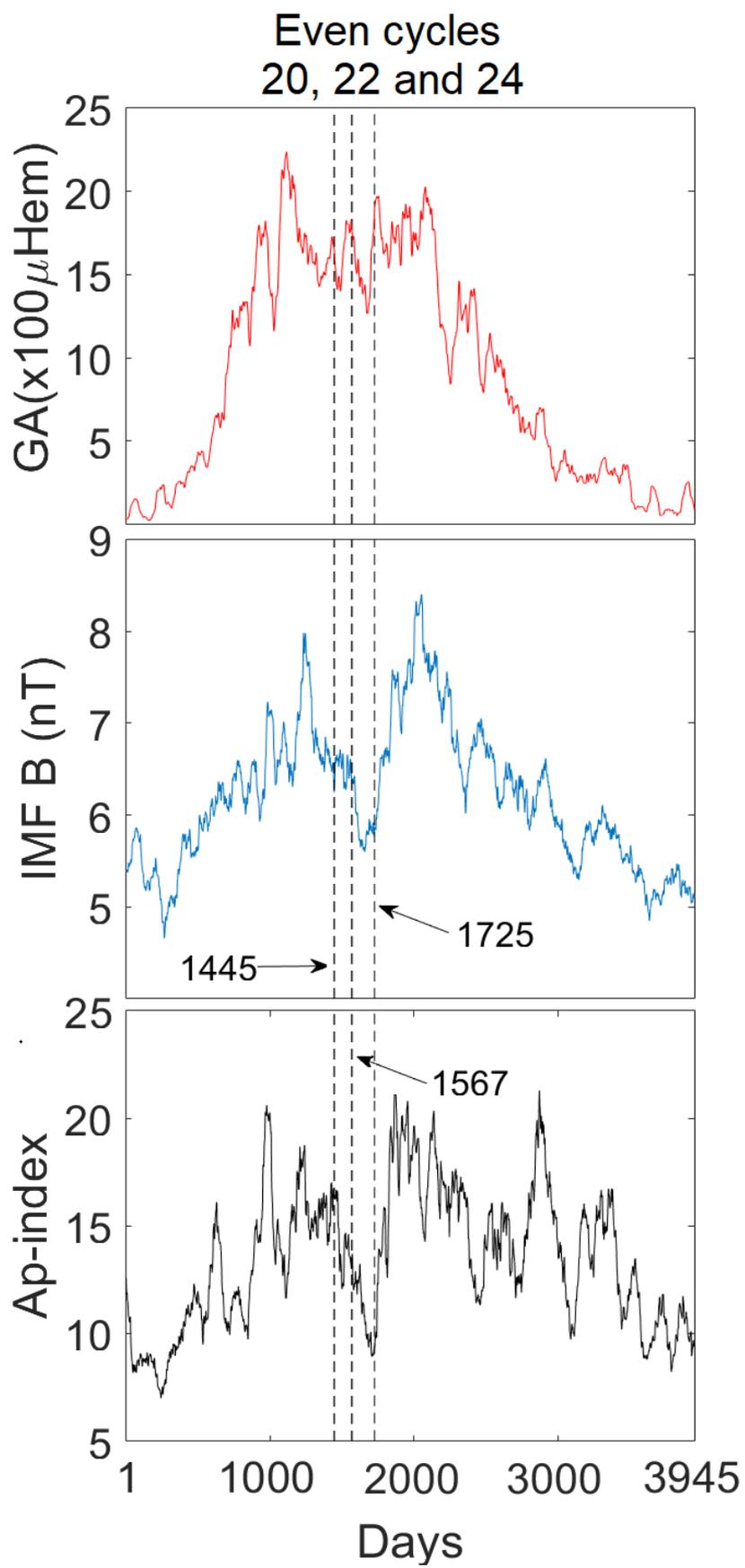

**Figure 12.**